# High temperature electronic behavior of $La_{0.8}Sr_{0.2}MnO_3$ thin film

G.T.Tan, S.Dai, P.Duan, H.B.Lu, K.Xie, Y.L.Zhou , Z.H.Chen*

Laboratory of Optical Physics, Institute of Physics & Center for Condensed Matter Physics, Chinese Academy of Sciences, P. O. BOX 603, Beijing 100080, P. R. China

**Abstract**

The electronic structure of $La_{0.8}Sr_{0..2}MnO_3$/SrTiO thin film, which was prepared by Laser MBE, was studied by X-ray photoemission spectra (XPS) in the temperature interval of 300 to 1000 K. Experimental results showed that the electronic state of the thin film underwent a discontinuous variation between 350 K and 450 K, indicating that the metal-semiconductor transition was probably a discontinuous phase transition. At high temperature (450 to 1000 K), both the binding energies of the atomic core level and valence-band of the film shifted up with increasing the temperature, while their line-widths became narrower, which were different from that observed at low temperature. These phenomena are attributed to the crystal lattice expansion and related Jahn-Teller distortion varying with temperature.

* Correspondence should be addressed to: Prof. Z.H. Chen, Laboratory of Optical Physics, Institute of Physics & Center for Condensed Matter Physics, Chinese Academy of Sciences, P.O. BOX 603, Beijing 100080, P.R. China.

* Email address: zhchen@aphy.iphy.ac.cn

## I. INTRODUCTION

For the past ten years, numerous studies have been focused on the hole-doped manganese perovskites compounds, $La_{1-x}D_xMnO_3$ (D= Ca, Sr, Ba and Pb ), which show colossal-magnetoresistance (CMR) behavior and related phase transitions[1-4] at low temperature. These phenomena were traditionally understood based on the double-exchange (DE) model and Jahn-Teller effects[5-7]. According to the model, these compounds were considered as in a mixed-valence system of $Mn^{3+}$-$Mn^{4+}$. The $Mn^{3+}$ ion has three $t_{2g}$ and one $e_g$ electron, the low energy $t^3_{2g}$ triplet state contributes a local spin of S=3/2, and the $e^1_g$ state electron is either itinerant or local depending on the spin related orientation. The hopping of $e^1_g$ electron between $Mn^{3+}$ and $Mn^{4+}$ and the dependence of hopping probability on the applied magnetic field qualitatively described the magneto resistance behavior. In these compounds, the electric transport and related physical phenomena are governed by the electronic structure, which depends on the average ionic radius of the perovskite A site and the degree of the hybridization between the $e^1_g$ state and O-2p state, and is also the function of temperature, pressure, and magnetic field.[8] Among of these compounds, $La_{1-x}Sr_xMnO_3$ plays an important role due to its higher transitional temperature and unusual physical behavior. Its electric structure were studied in detail by Chainani et al[9] and Saitoh et al[10] using electron spectroscopy techniques. However, these experiments were restricted to low temperature. In this paper, we report X-ray photoemission spectrum (XPS) studies on the electronic state and phase transition of $La_{0.8}Sr_{0.2}MnO_3$ thin film in high temperature region, namely 300–1000 K. The experimental data showed that the binding energies of the core level and valence band increased with temperature, while their line-widths narrowed. At the phase transition temperature, the electronic structure of the film underwent a dramatic variation. These unusual electronic behaviors of $La_{0.8}Sr_{0.2}MnO_3$ film could be attributed to the crystal lattice parameters and related Jahn-Teller distortion varying with temperature.

## II. EXPERIMENT

$La_{0.8}Sr_{0.2}MnO_3$ compound was prepared by traditionally solid-state reaction technique. The stoichiometric mixture of high purity $La_2O_3$, $SrCO_3$ and $MnO_2$ powders was pre-reacted at 800 $^{o}$C, following by sintering at temperature 1000 $^{o}$C and 1100 $^{o}$C with middle ground, and eventually it was sintered at 1300 $^{o}$C for about 24 hours in air. Its structure and single phase were checked by X-ray diffraction. Then it was used as a target to prepare the thin films of $La_{0.8}Sr_{0.2}MnO_3$. The thin film of $La_{0.8}Sr_{0.2}MnO_3$ with the thickness of 340 nm was grown one atomic layer by one atomic layer, on a (100) $SrTiO_3$ substrate using a laser molecular beam epitaxy (LMBE) system. A computer controlling system coupled with an in-situ reflection high-energy electron diffraction (RHEED) monitor system was used to control and monitor the growth process of the films. During the film growth, the substrate temperature was at about 670 $^{o}$C and the oxygen pressure was maintained under $2x10^{-1}$ Pa. The energy density and pulsed repetition rate of the laser source were about 2 J/cm$^2$ and 2 Hz, respectively. The details of the prepared technique are similar to those used in the fabrication of other perovskite oxidesprevious reported.[11] A good quality surface of the film has been obtained. The atomic force microscopy (AFM) image of the film revealed that the root mean square roughness (rms) of the surface was about 0.15 nm. X-ray diffraction (XRD) diagram indicated that the film had single perovskite phase structure with (001) orientation. The electric resistance was measured by the traditional four-probe method in the temperature range of 5 to 380 K. The temperature curve of the electric resistance showed that the temperature of the metal-semiconductor transition $T_{MS}$ was about 366 K. our $T_{MI}$ was closed to the value reported by Urushibara et al,[12] in which the Curie temperature Tc and the $T_{MI}$ of $La_{0.8}Sr_{0.2}MnO_3$ were determined to be 309 K and 360 K, respectively. In order toinvestigate the electronic state and the phase transitions of $La_{0.8}Sr_{0.2}MnO_3$ film at high temperature, the measurement of XPS on $La_{0.8}Sr_{0.2}MnO_3$ film was carried out on a VG photoelectron spectrometer with Mg Kα source. Before measuring, the surface of sample was cleared by Argon ionic etching and a subsequent ditto heat treatment was performed at about 800 $^{o}$C for about 30 minutes. The binding energies (B.E.) of all spectra were calibrated against the C-1s peaks. The C-1s peaks were at about 284.6, 285.02, 285.3, 284.98 and 284.78 eV at 300, 350, 450, 800 and 1000 K, respectively. To analyze and compare the obtained experiment data, the background of spectra due to the secondary electrons was subtracted and the intensity was normalized to integrated intensity. Finally, all the spectra were fitted with Gauss function.

## III. RESULTS AND DISCUSSIONS

Fig. 1(a) shows the XPS spectra of Mn-2p level of $La_{0.8}Sr_{0.2}MnO_3$ thin film at 350 K and 450 K. The XPS pattern displays a relative shift and broadening due to the phase transition. The binding energy of Mn-$2p_{3/2}$ is at about 641.47 eV and 641.15 eV for 350 K and 450 K, respectively. As the temperature rises from 350 K to 450 K, the binding energy shifts about –0.32 eV and the line-width (FWHM) widens about 0.15 eV. Fig. 1(a) reveals that the binding energy of Mn-2p electron has an abrupt reduction, although it exists the compensation effect due to temperature increasing. In the same interval of temperature, the similar phenomena are observed in the La-3d and Sr-3d electron state as well, which are shown in Fig. 1(b),1(c) and the insets. Previous reports have discussed the phase diagram and related phase transitions of $La_{1-x}Sr_xMnO_3$, which are the function of doped content, pressure and magnetic field;[8] however, the physical nature of the transitions is far from clear. In general, the phase transition, which occurred at relatively low temperature, was considered as a first order transition, while the transition taking place at a higher temperature was viewed as a continuous or a second order transition.[13] However, in our experiments, the binding energy of Mn-2p and other core level in $La_{0.8}Sr_{0.2}MnO_3$ displays an abrupt reduction as the temperature rises from 350 to 450 K. In this temperature interval, two events have happened: the temperature changed and the metal-semiconductor transition at 366 K. The temperature effect ought to induce the binding energy increasing with temperature (see the next paragraph). Thus, the discontinuous change could be considered as a result of the transition. The transition that undergoes at high temperature is probably a discontinuous or a first order phase transition. This is consistent with Radaelli's results[14] that the structure parameters underwent a discontinue transition at the metal-semiconductor transition temperature. Moreover, the line-width of the Mn-2p level increases when the temperature goes up from 350 to 450 K, and then decreases as the temperature increases continuously. Therefore the anomalous phenomena observed in XPS probably originate from the metal-semiconductor phase transition.

Fig. 2(a) is the Mn-2p core-level spectra of $La_{0.8}Sr_{0.2}MnO_3$ thin film at 450 K and 800 K. As the temperature rises from 450 K to 800 K, the Mn-2p level shifts about 1.09 eV and the line-width narrows about 330 meV. Fig. 2(b) exhibits the temperature dependence of the binding energy and line-width of Mn-2p level, which shows that the binding energy increases with increasing the temperature, regardless whether the sample is in the metal region or in the semiconductor one. For the semiconductor state, the temperature curve of B.E. can be fitted with an exponential function $E_{B.E}$=641+0.01952exp(0.00416T). The temperature effect is a positive one. The thermal shift of the core-level binding energy and the thermal narrowing or broadening of the bandwidth have been observed in semiconductor, alkali metal, 3d transition metal, 5d metal and oxides. Generally, its interpretation is based on the model of Hedin and Lundqvist,[15] that is, the thermal effect was attributed to the shift of the Fermi level, the changes of the electrostatic potential and the relaxation energy with temperature.

On the other hand, Fig. 2b displays that the line-width of the Mn-2p level of the $La_{0.8}Sr_{0.2}MnO_3$ decreases with increasing temperature, which is different from that in semiconductor and metal. Previous reports pointed out that the bandwidth of the core level was broadened with increasing the temperature due to the phonon broadening effect,[12,16-18] although there existed the temperature compensation induced by the lattice dilatation.[19] In contrast, the line-width of the Mn-2p level of the film is narrowed as the temperature increasing, which is probably due to the competition between lattice expansion and phonon broadening. In $ABO_3$ perovskite compounds, the bent B-O-B bond angle gradually straightens out and the B-O bond length elongates as the temperature increasing. The widening of the bond angle induces the increase of bandwidth, which is not very sensitive to the temperature variation. On the other hand, the dilatation of the bond length reduces the linewidth and is sensitive to the temperature change.[14] This suggests that the lattice expansion and related Jahn-Teller distortions varying with temperature have an important influence on the bandwidth of core level. In addition, the lifetime broadening of hole and the instrumental contribution has been considered as a constant.

As mentioned above, when the temperature increasing, the temperature drives the binding energy of the atomic core level to shift up and line-width to narrow, while the phase transition drives them to shift down and widen. Besides, the temperature also induces the change of the valence electronic state, as the temperature rises from room temperature to 1000 K. Fig. 3(a) is the valence band spectra of $La_{0.8}Sr_{0.2}MnO_3$ thin film at room temperature and 350 K. The spectral line shape with two peaks structure are similar to those observed by Chainani et al, [9] that was attributed to the nonbonding state and bonding state of the hybridization between Mn-3d and O-2p electrons. When the temperature increases to 350 K, the fine structure gradually disappears, but the bandwidth and Fermi level do not show apparent variation, although a slight shift of Mn-2p level has been observed. It is plausible that the valence band and core level are insensitive to temperature in the metallic state, which is similar to the results observed by T.Saith et al[10]. On the other hand, in the semiconductor state of high temperature, the line-width and the valence-band edge are very sensitive to temperature and show significant variance with increasing temperature. As shown in Fig. 3(b), in the temperature range of 450 to 800 K, the Fermi level, defined as the middle point of the valence band edge, shifts back about 1.5 eV and the line-width narrows about 1.3 eV, while the shift of the band bottom is small, which are different from the results what observed at the low temperture.[18] The binding energy and line-width of the valence band are a continuous function of temperature, and it could have originated from the similar mechanism that observed in the core level, which is due to the lattice expansion and related Jahn-Taller distortion, and due to the fact that the phonon broadening is not enough to compensate the lattice dilation effect. At the phase transition temperature, the Fermi level shift and the line-width change have been observed as well. As mentioned above, the film is in the metallic state at 350 K and in the semiconductor state at 450 K. The Fermi level of the semiconductor state relative to that of the metallic state, shifts towards lower binding energy side, and the line-width broadening is about 800 meV.

## IV. CONCLUSION

In summary, the XPS measurement on the $La_{0.8}Sr_{0.2}MnO_3$ thin film has been carried out at high temperature. The spectra revealed that the binding energy and line-width of the core level as well as valence band had a dramatic change at the metal-semiconductor transition, and the binding energy rose continuously with increasing temperature at high temperature region. Our experiment data showed that the transition taking place at high temperature is a discontinuous transition. The temperature effect on the electronic structure of the $La_{0.8}Sr_{0.2}MnO_3$ thin film could be stemmed from the thermal effect of the lattice expansion and related Jahn-Teller distortion. The temperature behavior of the electronic structure in the film was also different from that of a usual semiconductor and pure metal, although the temperature dependence of its resistance was similar to one of a semiconductor at high temperature. Thus, $La_{0.8}Sr_{0.2}MnO_3$ thin film can not be considered as a typical semiconductor.

**ACKNOWLEDGMENTS**

The author G.T. Tan would like to thank Kan Xie for her help in XPS measurements. This work was supported by a grant for State key Program No. G1998061412 of China.

# Figure Captions

Fig.1. XPS spectra of $La_{0.8}Sr_{0.2}MnO_3$ thin film at the temperature 350 K and 450 K: (a). Mn-2p, (b). Sr-3d. (c). La-$3d_{5/2}$. The inset shows the Sr-3d and La-$3d_{5/2}$ core level XPS spectra of the film at 450 K and 800 K, respectively.

F.g. 2. (a) XPS spectra of the Mn2p level of $La_{0.8}Sr_{0.2}MnO_3$ thin film at the temperature 450 K and 800 K, (b) the temperature dependence of Mn2p binding energy and FWHM for $La_{0.8}Sr_{0.2}MnO_3$ thin film, the fitting curve for the binding energy is shown in solid line. M = metal, S = semiconductor.

Fig.3. The valence band spectra of $La_{0.8}Sr_{0.2}MnO_3$ thin film (a) at R.T. and 350 K, (b) at 450 K and 800 K.

Fig.4. The valence band spectra of $La_{0.8}Sr_{0.2}MnO_3$ thin film at 350 K and 450 K

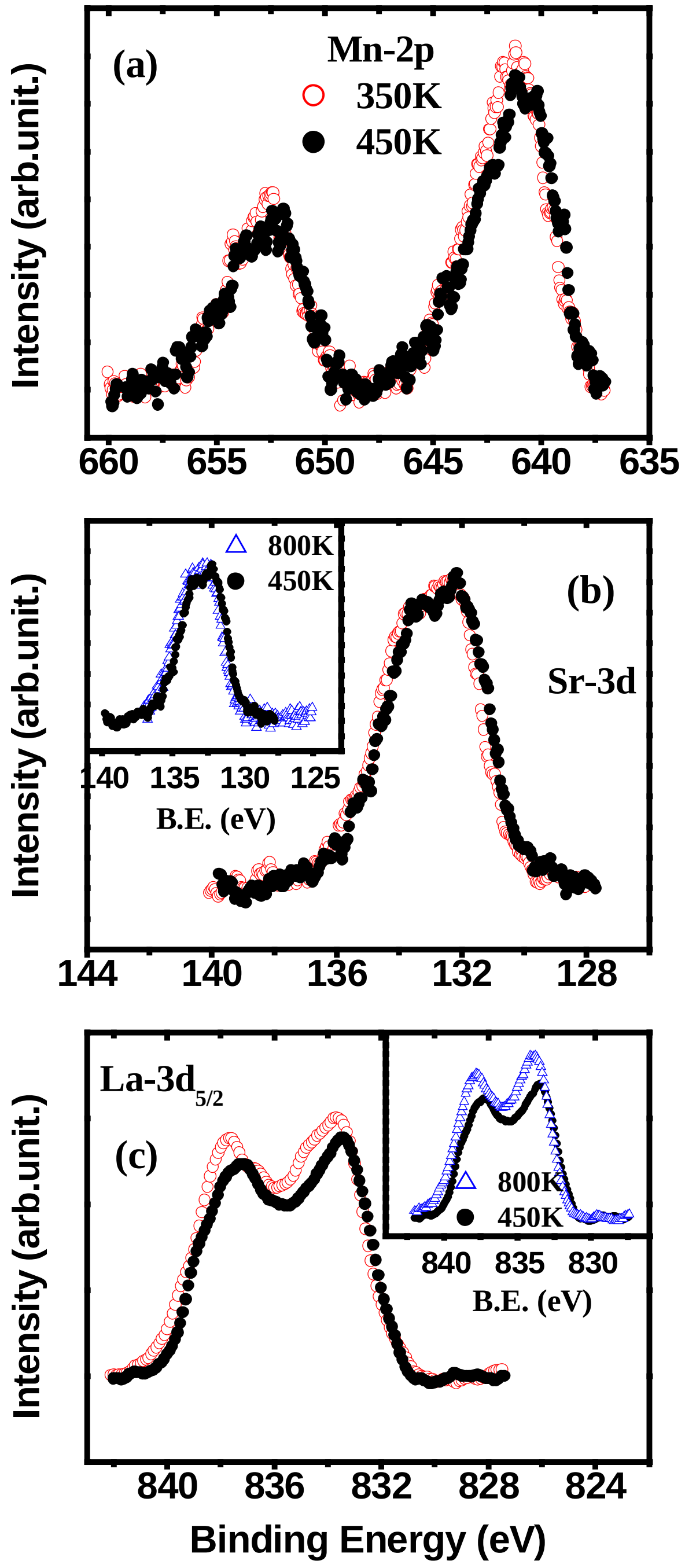

Fig. 1. G.T.Tan et al submitted to Phys.Rev.B

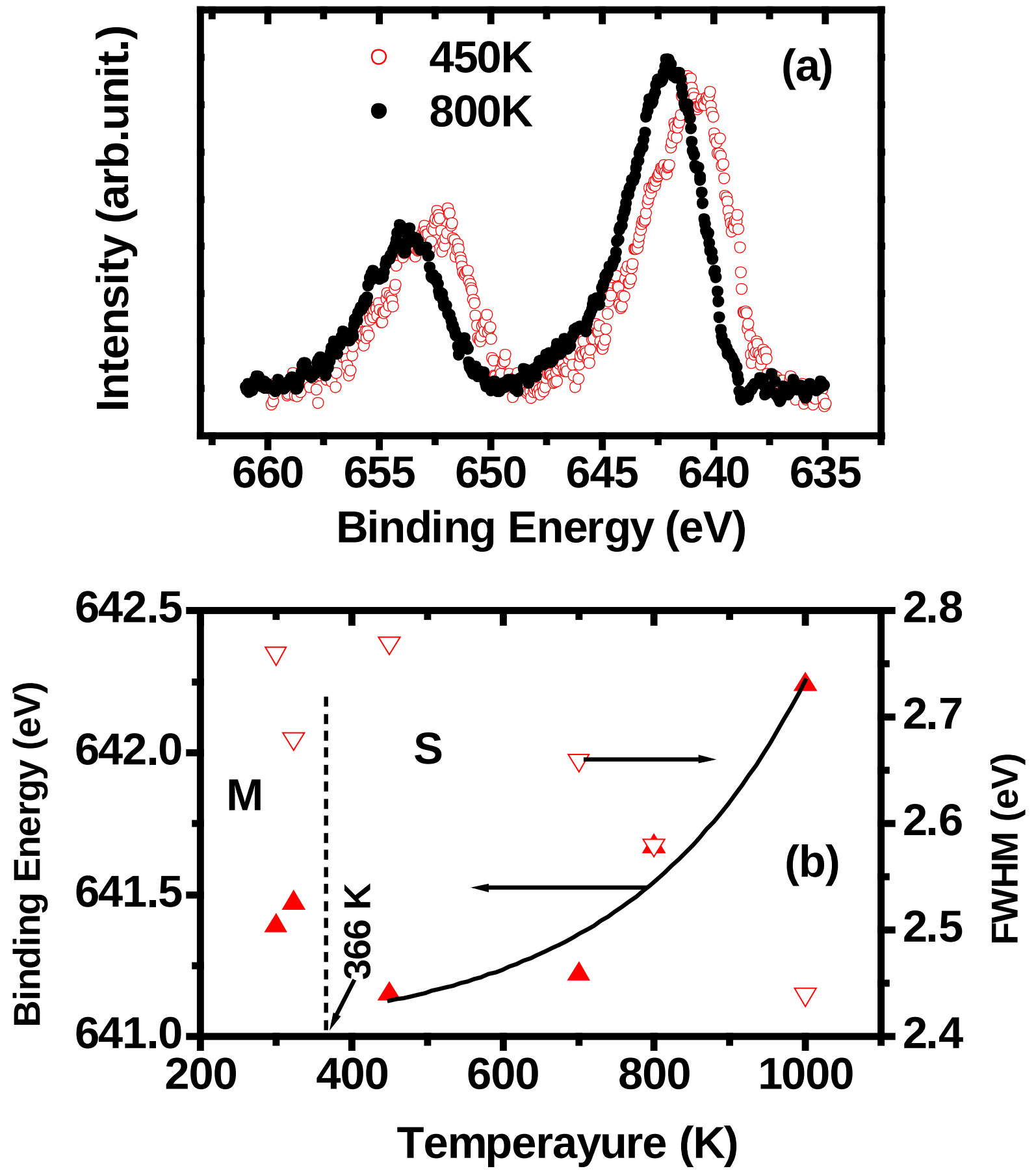

Fig. 2. G.T.Tan et al submitted to Phys.Rev.B

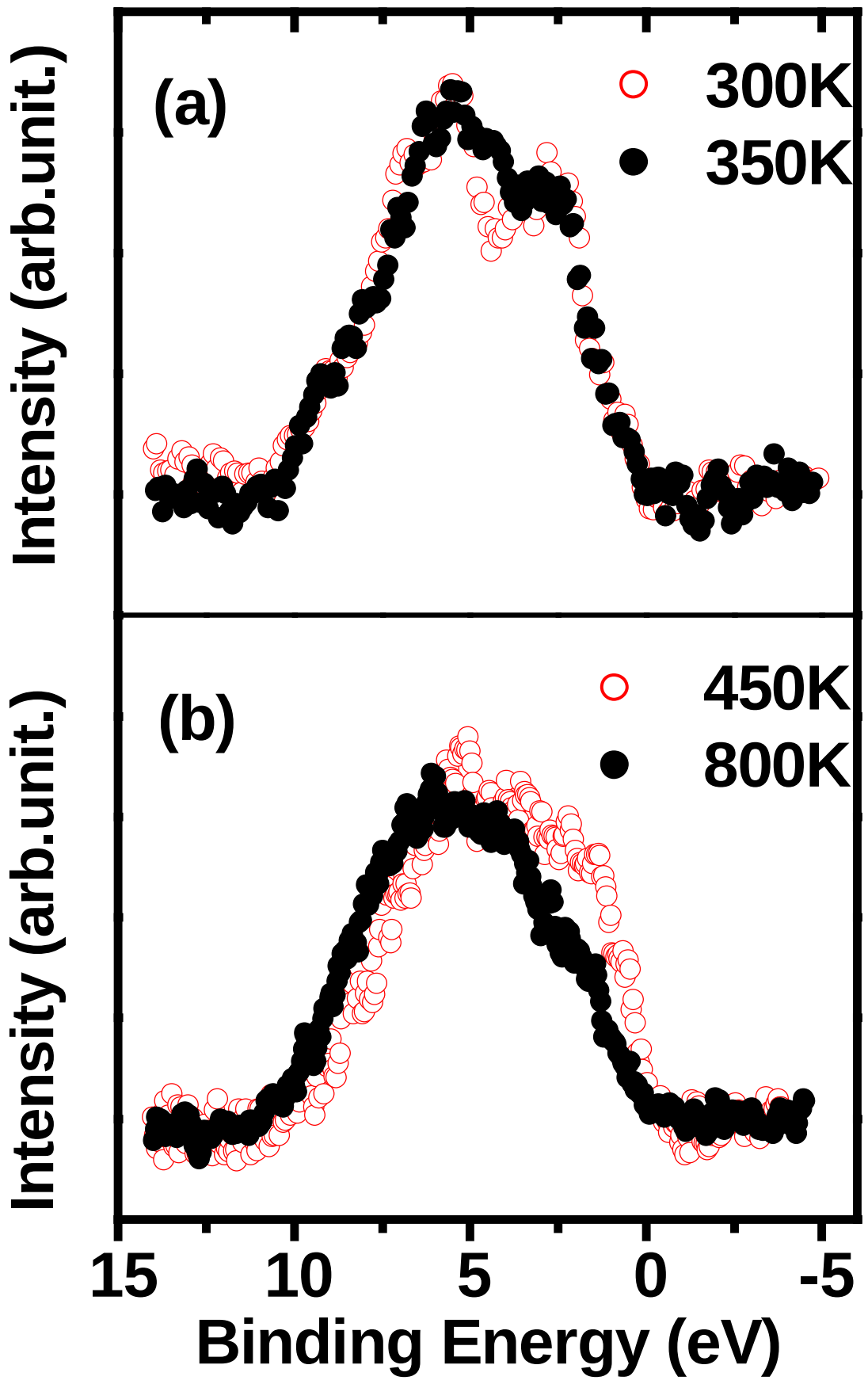

**Fig. 3. G.T.Tan et al submitted to Phys.Rev.B**

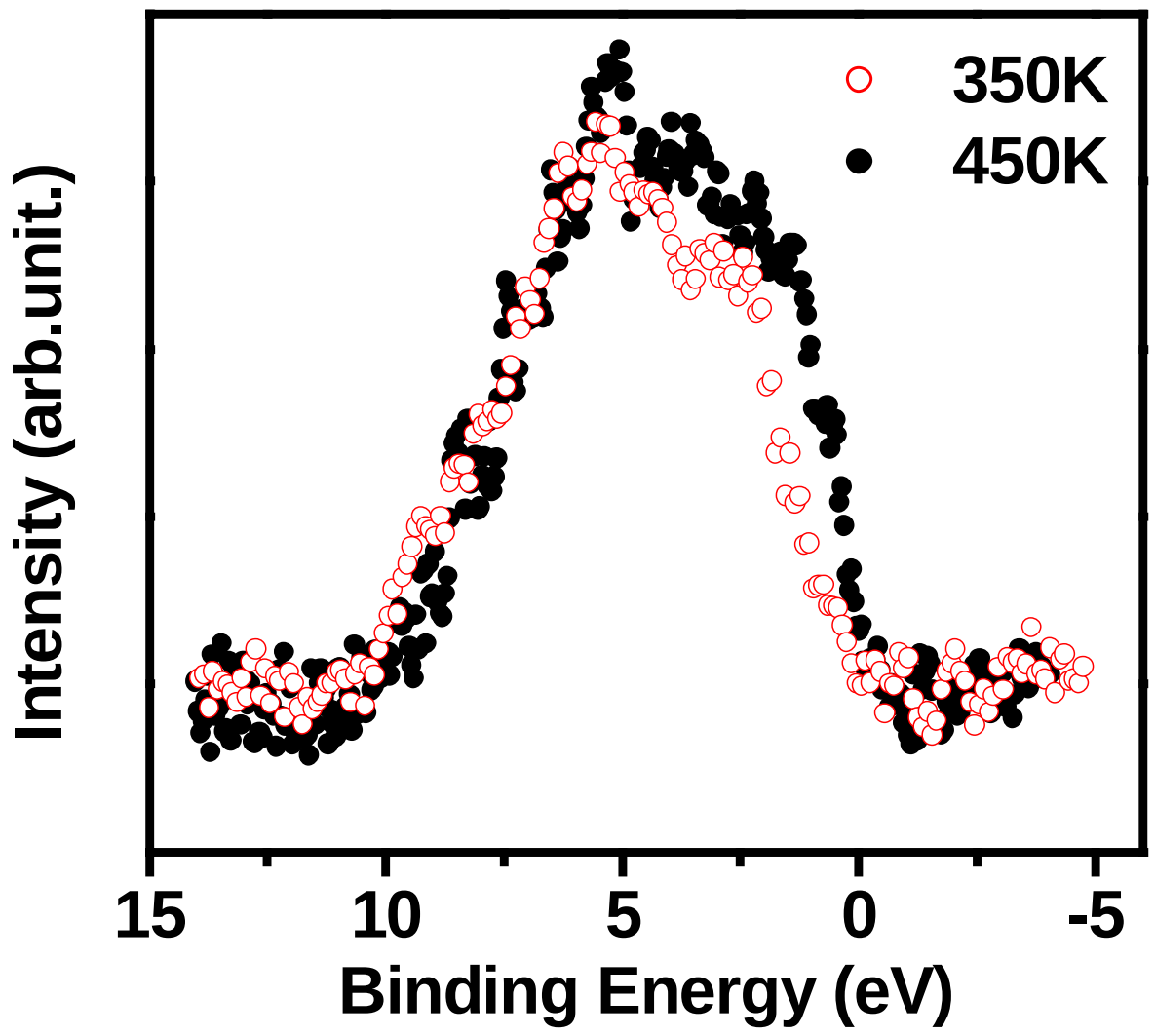

**Fig. 4. G.T.Tan et al submitted to Phys.Rev.B**